\documentclass[AMA,STIX1COL]{WileyNJD-v2}
\usepackage[utf8]{inputenc}
\def\doiheadtext{\relax}
\usepackage{tikz}

\usetikzlibrary{chains,arrows,automata,decorations.markings,positioning,calc,decorations.pathreplacing,patterns}
\usepackage{paralist}
\usepackage{subfloat,subfig}
\usepackage{siunitx}
\usepackage{underscore}
\usepackage{fixltx2e} % for subscript
\date{June 2019}
\usepackage{url}
\usepackage[inline]{enumitem}
 \usepackage{longtable}
\usepackage[frozencache,cachedir=.]{minted}
\newminted{cpp}{breaklines}
\usepackage{todonotes}
\DeclareMathOperator{\bitwiseand}{bitand}
\DeclareMathOperator{\bitwiseor}{bitor}
\DeclareMathOperator{\bitwisexor}{xor}

\usepackage{multirow}
\usepackage{graphicx}

\articletype{Research Article}%

\makeatletter

\newif\ifmintedminipage

\def\getoptarg[#1]{%
  \endgroup
  \ifthenelse{\equal{#1}{}}%
    {\mintedminipagefalse\begin{cppcode}}%
    {\mintedminipagetrue
      \begin{minipage}{#1}\begin{cppcode}}%
}

\newenvironment{cpp}%
  {\VerbatimEnvironment
    \begingroup\obeylines
    \@ifnextchar[{\getoptarg}{\getoptarg[]}}%
  {\end{cppcode}%
    \ifmintedminipage\end{minipage}\fi
  }

\newenvironment{cpp*}[2][]%
  {\VerbatimEnvironment
    \ifthenelse{\equal{#1}{}}%
      {\mintedminipagefalse\begin{cppcode*}{#2}}%
      {\mintedminipagetrue
        \begin{minipage}{#1}\begin{cppcode*}{#2}}%
  }%
  {\end{cppcode*}%
    \ifmintedminipage\end{minipage}\fi
  }

\makeatother
\begin{document}

\title{Efficient Computation of Positional Population Counts Using SIMD Instructions}
\author[1,2]{Marcus D. R. Klarqvist}
\author[3]{Wojciech Muła}
\author[4]{Daniel Lemire*}

\authormark{Klarqvist, Muła and Lemire}
\address[1]{\orgdiv{Department of Genetics}, \orgname{University of Cambridge}, \orgaddress{Downing Street, Cambridge CB2 3EH , \country{United Kingdom}}}
\address[2]{\orgname{Wellcome Sanger Institute}, \orgaddress{Wellcome Genome Campus, Hinxton, Cambridge, CB10 1SA, \country{United Kingdom}}}
\address[3]{0x80.pl, \orgaddress{\state{Wrocław}, \country{Poland}}}
\address[4]{\orgdiv{TELUQ}, \orgname{Universit\'e du Qu\'ebec}, \orgaddress{\state{Quebec}, \country{Canada}}}

\corres{*D. Lemire, Universit\'e du Qu\'ebec (TELUQ), 5800, Saint-Denis street, Montreal (Quebec) H2S 3L5, Canada. \email{daniel.lemire@teluq.ca}}

\abstract[Summary]{

In several fields such as statistics, machine learning, and bioinformatics, categorical variables are frequently represented as one-hot encoded vectors. 
For example, given 8~distinct values, we map each value to a byte where only a single bit has been set. We are motivated to quickly compute statistics over such encodings.

Given a stream of $k$-bit words, we seek to compute $k$~distinct sums corresponding to bit values at indexes 0, 1, 2, \ldots, $k-1$. If the $k$-bit words are one-hot encoded then the sums correspond to a frequency histogram.

This multiple-sum problem is a generalization of the population-count 
problem where we seek the sum of all bit values. % in a $k$-bit word.
Accordingly, we refer to the multiple-sum problem as a \emph{positional population-count}. 

Using SIMD (Single Instruction, Multiple Data) instructions from 
recent Intel processors, we describe algorithms for computing the 16-bit position population count using less than half
of a CPU cycle per 16-bit word. Our best approach uses up to 400~times fewer instructions and is up to 50~times faster than baseline code using only regular (non-SIMD) instructions, for sufficiently large inputs.
}

\maketitle

\section{Introduction}\thispagestyle{empty}

In many applications such as deep learning\cite{Dai2017,Courbariaux2016,Zhang2018,Hubara2016}, indexing\cite{Lemire2016}, chemistry\cite{Haque2011}, 
cryptography\cite{Sanyal2018}, and bioinformatics\cite{Layer2016,Danek2018,Wu2010,Purcell2007} it is desirable to compute the number of set bits in a computer word. 
This operation is referred to as the population count (\texttt{popcnt}), Hamming weight, or the sideways sum of the word. 
For example, the machine word \texttt{\underline{1}00\underline{1}00\underline{1}0} has a population count of three since there are three set bits. We have previously described efficient subroutines for computing the population count\cite{mula2017faster} of large arrays that
takes advantage of SIMD (Single Instruction, Multiple Data) instructions available on most commodity
processors. These instructions operate on wide data registers and reduce the number of instructions required by processing multiple machine words simultaneously.

It is common to represent categorical variables using one-hot (1-of-$k$) encoding\cite{lippert2013exhaustive,mittag2015influence} where each categorical value maps to a corresponding bit within a $k$-bit word and each word may only have a single bit set.
As an illustrative example, consider a categorical variable \texttt{country} that has the eight distinct values \texttt{Australia}, \texttt{Canada}, \texttt{China}, \texttt{France}, \texttt{Japan}, \texttt{Portugal}, \texttt{Spain}, and \texttt{USA}. These categorical string values can be unambiguously dictionary-compressed into one-hot encodings (Table~\ref{table:one-hotexample}) resulting in lower memory usage and more efficient queries. One of the motivations for such categorical encodings is that many machine-learning algorithms require numerical variables and cannot operate directly on categorical data. These encodings are closely related to the concept of dummy variables in statistics where categorical attributes are represented as zeros or ones for computing purposes. In this context, we want to compute the frequency histogram, that is the number of occurrences of each value. For example, given a stream (Table~\ref{table:one-hotexample-stream}) of one-hot encoded values (Table~\ref{table:one-hotexample}), we want to compute the number of times each country has been observed such that \texttt{France} has been observed three times, \texttt{Portugal} two times, \texttt{USA} four times, and \texttt{China} one time.
To compute such cumulative frequency histograms from encoded words, we need to count the number of set bits at each position (first, second, \ldots, last).

\begin{table}
\caption{Example of a one-hot encoding.} \centering
\subfloat[mapping\label{table:one-hotexample}]{
\begin{tabular}{ll}
\toprule
Value & One-hot\\  \midrule
\texttt{Australia} &  \texttt{\underline{1}0000000} \\
\texttt{Canada}    &  \texttt{0\underline{1}000000} \\
\texttt{China}     &  \texttt{00\underline{1}00000} \\
\texttt{France}    &  \texttt{000\underline{1}0000} \\
\texttt{Japan}     &  \texttt{0000\underline{1}000} \\
\texttt{Portugal}  &  \texttt{00000\underline{1}00} \\
\texttt{Spain}     &  \texttt{000000\underline{1}0} \\
\texttt{USA}       &  \texttt{0000000\underline{1}} 
   \\ \bottomrule
\end{tabular}
}
\subfloat[Stream of values\label{table:one-hotexample-stream}]{
\begin{tabular}{l}
\texttt{000\underline{1}0000} (France)\\
\texttt{000\underline{1}0000} (France)\\
\texttt{00000\underline{1}00} (Portugal)\\
\texttt{000\underline{1}0000} (France)\\
\texttt{0000000\underline{1}} (USA)\\
\texttt{00000\underline{1}00} (Portugal)\\
\texttt{0000000\underline{1}} (USA)\\
\texttt{0000000\underline{1}} (USA)\\
\texttt{0000000\underline{1}} (USA)\\
\texttt{00\underline{1}00000} (China)
   \\ 
\end{tabular}
}
\end{table}

\begin{figure}
    \centering 
    \subfloat[\label{fig:introductionpane2}per-word population count]{\includegraphics[trim={0 0  1.6cm 0},clip,scale=1.5]{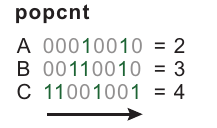}}\hspace*{2cm}
    \subfloat[\label{fig:introductionpane1}positional population count]{\includegraphics[trim={0 -0.2cm  0cm 0},clip,scale=1.5]{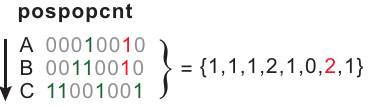}}\hspace{1cm}
%\subfloat[positional population count as a popul]{\includegraphics[trim={0 0 0 0cm},clip,scale=1.5]{pospopcnt-overview-pane3}}
    \caption{\label{fig:introduction} Comparing per-word population count (\texttt{popcnt}) with positional population count (\texttt{pospopcnt}).}
\end{figure}

We name this novel generalization of the population count operation to individual bits spanning multiple words %in a novel columnar population count operation referred to 
as the \textit{positional population count} (\texttt{pospopcnt}). If each word is made of $k$-bits then we want to compute $k$~counts representing the total number of set bits at each of the $k$~positions. As an illustrative example, consider three given input words $A$, $B$, and $C$ (Fig.~\ref{fig:introduction}), the conventional per-word population count computes the number of set bits for each word independently (Fig.~\ref{fig:introductionpane2}). In contrast, the positional population count operation computes the \emph{vertical} population count over these words for independent bit positions (Fig.~\ref{fig:introductionpane1}).

Our main contribution is the formalized description of the novel positional population count operation and the introduction of an efficient algorithmic family for computing it using SIMD instructions leveraging the AVX-512 instruction set architecture (ISA) available on recent Intel processors. The AVX-512 family of ISAs can operate on up to thirty-two 512-bit registers.
%they  operate on registers as large as 512~bits. %They are an upgrade from the older SSE 
Notably, we show that our proposed algorithm can be in excess of 50-fold faster compared to a baseline scalar approach, as long as the input is sufficiently large (several kilobytes).

\section{Related Work} \label{sec:relatedwork}

To our knowledge, there is no prior work in the literature describing the positional population count. In contrast, computation of the population count has enjoyed renewed interest\cite{mula2017faster} following recent advancements in hardware with wider registers and more diverse and mature instruction-set architectures. Computing the population count is of such general importance that most commodity processors have dedicated instructions (e.g. \texttt{popcnt} on x64 processors). These hardware instructions are  efficient when operating on single words. When large volumes of machine words are available in contiguous memory, other software-based approaches have demonstrated superior performance\cite{mula2017faster}. 

One of the earliest efficient algorithm to compute the population count is the Wilkes-Wheeler-Gill population-count\cite{Wilkes} (Fig.~\ref{fig:tree-of-adders}). This algorithm begins by adding adjacent bits into two-bit subwords, then adds these two-bit subwords into four-bit subwords, and finally adds adjacent four-bit subwords into byte values. These byte values are then summed using a multiplication and a bit-shift. It has a natural tree structure and can be described as a "tree of adders"\cite{KnuthV4A}.

\begin{figure}
\centering
%%%%%%%%%%%%%%%%%%%%%%%%%%%%%%%%%%%%%%%%
\begin{tikzpicture}[level distance=1.5cm,
  level 1/.style={sibling distance=8cm},
  level 2/.style={sibling distance=4cm},
  level 3/.style={sibling distance=2cm}]

  \node[fill=green!10,draw] {population count of 8~bits}
  child {node[fill=red!10,draw] {sum of bits 1, 2, 3, 4}
    child {node[fill=blue!10,draw] {sum of bits 1 and 2}
      child {node[fill=yellow!30,draw]  {1$^{\text{st}}$ bit}  edge from parent[<-,thick,>=latex]}
      child {node[fill=yellow!30,draw] {2$^{\text{nd}}$ bit} edge from parent[<-,thick,>=latex]}
      edge from parent[<-,thick,>=latex]
    }
    child {node[fill=blue!10,draw] {sum of bits 3 and 4}
    child {node[fill=yellow!30,draw] {3$^{\text{rd}}$ bit} edge from parent[<-,thick,>=latex]}
      child {node[fill=yellow!30,draw] {4$^{\text{th}}$ bit} edge from parent[<-,thick,>=latex]}
      edge from parent[<-,thick,>=latex] 
    }
    }
    child {node[fill=red!10,draw] {sum of bits  5, 6, 7, 8}
    child {node[fill=blue!10,draw] {sum of bits  5 and 6 }
      child {node[fill=yellow!30,draw]  {5$^{\text{th}}$ bit}  edge from parent[<-,thick,>=latex]}
      child {node[fill=yellow!30,draw] {6$^{\text{nd}}$ bit} edge from parent[<-,thick,>=latex]}
      edge from parent[<-,thick,>=latex]
    }
    child {node[fill=blue!10,draw] {sum of bits 7 and 8}
    child {node[fill=yellow!30,draw] {7$^{\text{rd}}$ bit} edge from parent[<-,thick,>=latex]}
      child {node[fill=yellow!30,draw] {8$^{\text{th}}$ bit} edge from parent[<-,thick,>=latex]}
      edge from parent[<-,thick,>=latex] 
    }
    };
\end{tikzpicture}
%%%%%%%%%%%%%%%%
\caption{\label{fig:tree-of-adders} A tree of adders to compute the population count of eight bits in three steps.}\end{figure}
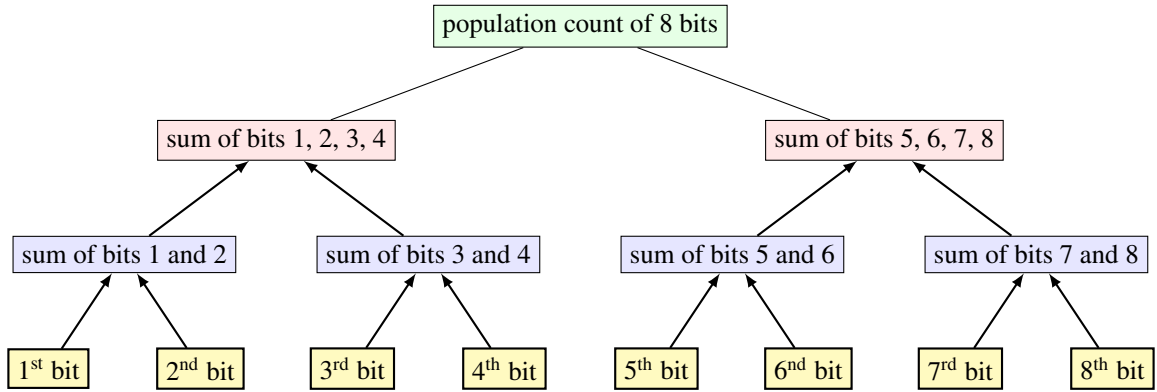

\begin{figure}\centering
\begin{cpp}[0.6\columnwidth]
void CSA(uint64_t* h, uint64_t* l, 
    uint64_t a, uint64_t b, uint64_t c) {
  uint64_t u = a xor b;
  *h = (a bitand b) bitor (u bitand c);
  *l = u xor c;
}
\end{cpp}
\caption{\label{fig:csacpp}A C++ function  implementing a bitwise parallel carry-save adder (CSA). Given three input words $a,b,c$, we compute $h,l$ representing the most significant and the least significant bits of the sum of the bits from $a$, $b$, and $c$, respectively.}
\end{figure}

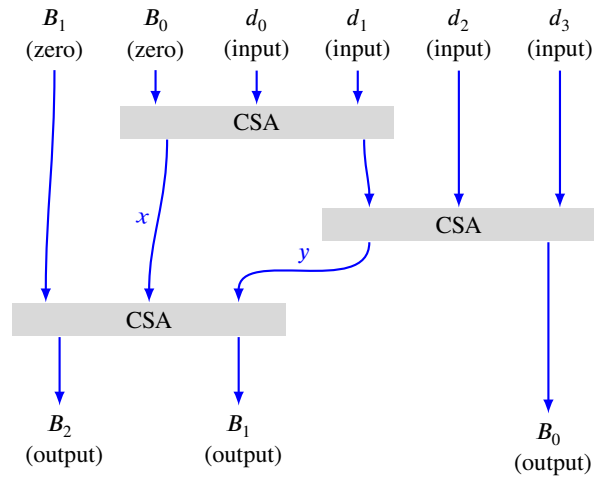
\begin{figure}
\centering
%\parbox{0.8\textwidth}{%
\begin{tikzpicture}[thick,scale=0.9, every node/.style={transform shape},node distance=0cm,start chain=1 going right,
start chain=9 going right,start chain=10 going right,start chain=11 going right,start chain=12 going right,start chain=13 going right]
 \tikzstyle{mytape}=[minimum height=0.7cm,minimum width=1.45cm]

 %   \node(A1) [on chain=9,mytape] {sixteens};
%    \node(A2) [on chain=9,mytape] {\texttt{eights}};
%    \node(A3) [on chain=9,mytape] {\ldots};
    \node(A4) [on chain=9,mytape] {\parbox{1cm}{\centering $B_1$\\(zero)}};
    \node(A5) [on chain=9,mytape] {\parbox{1cm}{\centering $B_0$\\(zero)}};
    \node(A6) [on chain=9,mytape] {\parbox{1cm}{\centering $d_0$\\(input)}};
    \node(A7) [on chain=9,mytape]
    {\parbox{1cm}{\centering $d_1$\\(input)}};
%%%%%%%%%%%

\node(csa1)[on chain=10,fill=gray!30,minimum width=4cm,below=0.5cm of A6] {CSA};
    \node(A8) [on chain=9,mytape]
    {\parbox{1cm}{\centering $d_2$\\(input)}};
    \node(A9) [on chain=9,mytape]
    {\parbox{1cm}{\centering $d_3$\\(input)}};
\draw[->,blue, >=latex,thick] (A5.south)  to    (A5.south |- csa1.north);
\draw[->,blue, >=latex,thick] (A6.south)  to    (A6.south |- csa1.north);
\draw[->,blue, >=latex,thick] (A7.south)  to    (A7.south |- csa1.north);
%%%%%%%%%%%
\node(csa2)[on chain=11,fill=gray!30,minimum width=4cm,below=2cm of A8] {CSA};
%  \node(A10) [on chain=11,mytape] {\ldots};

\draw[->, blue,>=latex,thick]
    ($(csa1.south east)!0.22!(csa1.south)$) to[out=270,in=90]
    ($(csa2.north west)!0.35!(csa2.north)$);
\draw[->, blue,>=latex,thick] (A8.south)  to    (A8.south |- csa2.north);
\draw[->, blue,>=latex,thick] (A9.south)  to    (A9.south |- csa2.north);

\node(csa3)[on chain=12,below=1cm of csa2,minimum width=4cm] {};
\node(csa12)[fill=gray!30,minimum width=4cm,left=0.5cm of csa3] {CSA};

\draw[->,blue, >=latex,thick]
    (A4.south) to[out=270,in=90] 
    ($(csa12.north west)!0.25!(csa12.north)$) ;

\node(foursoutput)  [on chain=13,mytape,below=1cm of $(csa12.south west)!0.35!(csa12.south)$,minimum width=2cm]
{\parbox{1cm}{\centering $B_2$\\(output)}};

\draw[->,blue,>=latex,thick]
    ($(csa12.south west)!0.35!(csa12.south)$) to[out=270,in=90] 
    (foursoutput.north) ;

\node(twosoutput)  [minimum width=2cm,below=1cm of $(csa12.south east)!0.35!(csa12.south)$]
{\parbox{1cm}{\centering  $B_1$\\(output)}};

\draw[->,blue, >=latex,thick] 
    ($(csa12.south east)!0.35!(csa12.south)$) to[out=270,in=90] 
    (twosoutput.north);

\node(onesoutput)  [mytape,minimum width=2cm,below=2.5cm of $(csa2.south east)!0.35!(csa2.south)$]
{\parbox{1cm}{\centering  $B_0$\\(output)}};

\draw[->, blue,>=latex,thick]
    ($(csa2.south east)!0.35!(csa2.south)$) to[out=270,in=90]
    (onesoutput.north);

\draw[->,blue, >=latex,thick]
    ($(csa2.south west)!0.35!(csa2.south)$) to[out=270,in=90] node[midway,sloped,above] {$y$}
    ($(csa12.north east)!0.35!(csa12.north)$);

\draw[->,blue, >=latex,thick]
    ($(csa1.south west)!0.35!(csa1.south)$) to[out=270,in=90] node[midway,sloped,left,rotate=270] {$x$}
    (csa12.north)  ;
\end{tikzpicture}
%}
\caption{\label{fig:hsillustration}CSA circuit algorithm aggregating four inputs ($d_0, d_{1}, d_{2}$, $d_{3}$), producing three outputs $B_0$, $B_1$, $B_2$ corresponding to the least significant bit, second most significant bit and most significant bit of the sum. Observe that $B_0$ and $B_1$ are both inputs and outputs.}
\end{figure}

Another efficient approach\cite{warren2007} is the Harley-Seal algorithm that is based on carry-save adder (CSA) networks frequently used in hardware microarchitectures. The CSA subroutine takes three input words and produce two output words: (1) a sequence of partial sum bits, and (2) a sequence of carry bits.
Given three~bit values ($a,b,c\in \{0,1\}$), the sum $a + b + c$ is computed as a 2-bit word where the least significant bit is given by $(a\bitwisexor b) \bitwisexor c$ and the most significant bit is given by $(a \bitwiseand b) \bitwiseor ( (a \bitwisexor b) \bitwiseand c )$. We reuse the $a \bitwisexor b$ expression from the computation of the least significant bit when computing the most significant bit. Therefore, three values can be summed using 5~logical operations. This CSA network is particularly effective when generalized to operate on 64-bits in parallel (Fig.~\ref{fig:csacpp}).
Let us illustrate this approach by computing the population count of four inputs words using a simplified circuit (Fig.~\ref{fig:hsillustration}). Starting with two 64-bit words initialized to zero: one for the least significant bits ($B_0$) and one for the second least significant bits ($B_1$). 
\begin{enumerate}
    \item Two input words are summed with $B_0$ using a CSA subroutine (Fig.~\ref{fig:csacpp}) resulting in two outputs: one corresponding to the least significant bits stored in $B_0$ and another corresponding to the second most significant bit: let us call it $x$.
    \item Two new input words are summed with $B_0$ using the same CSA subroutine. Again the output corresponding to the least significant bit is stored in $B_0$. We have a second output corresponding to the second most significant bits, let us call it $y$.
    \item In the final step, $x$, $y$, and $B_1$ are added together. The output corresponding to the least significant bits is stored in $B_1$. Similarly, the output corresponding to the most significant bit is stored in $B_2$.
\end{enumerate}
If $c_i$ is the population count of $B_i$, then the population count of the four words is $1 c_0 + 2 c_1 + 4 c_2$. In general, given $2^N$~input words, we can produce $N+1$~words corresponding to the sum of the bits using a CSA circuit. Then the per-word population count can be computed from the resulting $N$~words. 
More complex circuits are difficult to illustrate but can be built by recursively applying the same circuit. %For example, if we had 8~inputs instead of 4~inputs, we might first apply the circuit from Fig.~\ref{fig:hsillustration} on the first 4~inputs, taking  $B_0$, $B_1$ as inputs (possibly initialized to zero) and  generating new values of $B_0$, $B_1$, $B_2$. We could then repeat the same process (circuit) with the last 4~inputs (out of 8), except that we should produce a different $B_2$, let us call it $B'_2$. As a final step, we would need to combine $B_2$, $B'_2$ to produce a new $B_2$ and a $B_3$ using a CSA routine (adding a zero input if needed). 
Assume that we have a circuit that takes $2^N$~inputs as well as  $B_0, \ldots, B_{N-1}$ and produces the updated $B_0, \ldots, B_{N}$ corresponding to the sum. We can represent this circuit as a function $F_N (d_0, \ldots, d_{2^N-1}; B_0, \ldots, B_{N-1})$ that outputs  $B_0, \ldots, B_{N}$. We can then construct a larger circuit that takes twice as many inputs by two calls to the circuit together with an additional CSA routine. Indeed, to compute $F_{N+1} (d_0, \ldots, d_{2^{N+1}-1}; B_0, \ldots, B_{N})$, we can start by computing $F_N (d_0, \ldots, d_{2^N-1}; B_0, \ldots, B_{N-1})$. We provide the first $N$~outputs ($B_0, \ldots, B_{N}$) from this first call to $F_N$ to a second call to the circuit $F_N$, this time using the $2^N$ following inputs $d_N, \ldots, d_{2^{N+1}-1}$. We are left with three values $B_N$, one was provided as an input to $F_{N+1}$, and two have been produced by a call to the smaller circuit $F_N$. We can combine these with a CSA routine. By such a recursive argument, we can show that $2^N-1$~CSA routines is sufficient to process $2^N$~inputs.

%\begin{figure}
%    \centering 
%    \includegraphics[width=0.5\textwidth]{diag1}
%    \caption{\label{fig:diag1} Construction of a wider CSA circuit by combining two existing circuits}
%\end{figure}

%We can fill these words with a circuit (Fig.~\ref{fig:csa-network}) and once the computation is completed, the total population count is the sum of the set bits in the first word plus two times the set bits in the second word, plus four times the set bits in the third word, and so forth. If generalized, 
In addition to being scaled to more inputs, the CSA circuit can be be parallelized using wide SIMD registers (e.g. 512-bit words using the AVX-512 ISA). Instead of processing 64-bit words one-by-one, we can process multiple 64-bit words in parallel resulting in greater speed. Given sufficiently large input arrays (e.g., 1\,kB), such a SIMD-accelerated CSA network is superior to all known population count alternatives on current Intel processors\cite{mula2017faster}. 

Furthermore, our general strategy and many of our constructions are familiar to cryptographers~\cite{chou2016qcbits}. Some of these researchers have produced related optimizations using advanced SIMD instructions~\cite{guimaraes2019optimized,drucker2020qc}.

\section{Carry-Save-Added (CSA) routines in AVX-512}

The conventional approach for computing the CSA routine requires 5~logical operations.
On processors with the AVX-512 instruction set architecture available, we can simplify this subroutine down to two instructions by using the three-operand instruction \texttt{vpternlogd} using the \texttt{\_mm512\_ternarylogic\_epi32} intrinsic function (Fig.~\ref{fig:csa-avx512}). These compiler-specific intrinsic functions enables access to specific processor instructions without having to directly write in assembly or machine language. The \texttt{vpternlogd} instruction operates on three input bits $a$, $b$, $c$ and an 8-bit sequence $i$ and returns the bit at index $1a + 2b + 4c$ of $i$. 
Given three 512-bit input registers, the instructions generates a 512-bit output, executing 512-bit operations in parallel. The \texttt{vpternlogd} instruction can compute any ternary Boolean function using a single instruction.

This instruction is useful in computing the most and least significant bits in the CSA routine. To determine the value of the 8-bit sequence $i$ in the context of computing the least and most significant bits, it suffices to enumerate all 8~possible inputs (Table~\ref{tab:csatern}).
\begin{itemize}
\item The least significant bit can be computed as the \texttt{xor} of the three inputs. Therefore, the  
 \texttt{vpternlogd} instruction computes the desired output when the bits at index $\{1, 2, 4, 7\}$ of $i$ are set. In software, this bit sequence is represented as the integer 0b10010110 in binary or 150 in decimal.
\item Computing the most significant bits involves the expression 
$(a \mathrm{~and~} b) \mathrm{~or~} ((a \mathrm{~xor~} b) \mathrm{~and~} c$.
We can achieve this result by setting the bits at index \{3, 5, 6, 7\} of $i$ that can be represented as the binary string 0b11101000 or 232 in decimal.
\end{itemize}

\begin{table}\centering
\caption{CSA routine in the context of the  \texttt{vpternlogd} instruction\label{tab:csatern}}
\subfloat[least significant bit]{
\begin{tabular}{ccccc}
\toprule
 $c$ & $b$& $a$ & $a \mathrm{~xor~} b \mathrm{~xor~} c$ & $1a + 2b + 4c$\\\midrule
 0& 0& 0& 0& 0\\
 1& 0& 0& 1& \textbf{1}\\
 0& 1& 0& 1& \textbf{2}\\
 1& 1& 0& 0& 3\\
 0& 0& 1& 1& \textbf{4}\\
 1& 0& 1& 0& 5\\
 0& 1& 1& 0& 6\\
 1& 1& 1& 1& \textbf{7}\\\bottomrule
\end{tabular}
}\hspace{1cm}
\subfloat[most significant bit]{
\begin{tabular}{ccccc}
\toprule
 $c$ & $b$& $a$ & $(a \mathrm{~and~} b) \mathrm{~or~} ((a \mathrm{~xor~} b) \mathrm{~and~} c)$ & $1a + 2b + 4c$\\\midrule
 0& 0& 0& 0& 0\\
 1& 0& 0& 0& 1\\
 0& 1& 0& 0& 2\\
 1& 1& 0& 1& \textbf{3}\\
 0& 0& 1& 0& 4\\
 1& 0& 1& 1& \textbf{5}\\
 0& 1& 1& 1& \textbf{6}\\
 1& 1& 1& 1& \textbf{7}\\\bottomrule
 \end{tabular}
}
\end{table}

\noindent When comparing the resulting AVX-512 routine to its 64-bit equivalent (Fig.~\ref{fig:csacpp}), we observe that the AVX-512-based CSA subroutine can process 8~times as many bits (512 versus 64) while simultaneously reducing the number of logical operations from 5 to 2. Taken together, we achieve a gain factor of $5/2 \times 8 = 20$ in terms of instructions per input bits.

\begin{figure}
\centering
\begin{cpp}[0.6\columnwidth]
void CSA_AVX512(__m512i* h, __m512i* l, 
   __m512i a, __m512i b, __m512i c)
{
  *h = _mm512_ternarylogic_epi32(c, b, a, 0b11101000);
  *l = _mm512_ternarylogic_epi32(c, b, a, 0b10010110);
}
\end{cpp}
\caption{\label{fig:csa-avx512} The carry-save adder update step using AVX-512-based instructions. The three inputs are $a, b, c$, the two outputs are $h$ and $l$ corresponding respectively to the most significant and least significant bits.}
\end{figure}

\section{Positional Population-Count Algorithms}\label{sec:algo}

Implementations of the positional population count depends on the desired target word size. For simplicity, we focus on 16-bit words in this work. Wider words (larger universes) can be encoded into multiple distinct streams of 16-bit words. In addition, our algorithms can be generalized to wider words without difficulty.

To compute the positional population count, we start with sixteen zero-initialized bit-counters. Next, we iterate over the bits in each word in sequence while incrementing the corresponding target bit-counter as needed. A sensible baseline algorithm for computing the positional population count involves  
a shift-mask-add subroutine (Fig.~\ref{fig:pospopcnt-naive}):
\begin{enumerate}
    \item Right shift a 16-bit word by the bit index $p\in\{ 0,1,\ldots,15\}$. Mathematically, this is equivalent to an integer division by $2^p$ with the result being the integer quotient.
    \item Mask out all but the least significant bit from the shifted word. This bit is set when the quotient is odd, and zero otherwise.
    \item Increment the counter at position $p$ with the value from step 2.
\end{enumerate}
We expect this code to be compiled to efficient assembly. On superscalar processors, branches must be predicted and a branch misprediction can cost tens of cycles or more. Fortunately, in our tests, the executed code does not trigger a significant number of mispredicted branches: the end of the short loop (16) is corrected predicted.

For sufficiently long input streams, we estimate the number of executed instructions to $\sim$4 per bit for a total of $\sim$64~instructions per 16-bit word. Given that most recent Intel processors can retire four instructions per CPU cycle, we expect a processing speed of around 16~cycles per 16-bit word for this scalar shift-mask-add subroutine (Fig.~\ref{fig:pospopcnt-naive}).

\begin{figure}\centering
\begin{cpp}[0.6\columnwidth]
void pospopcnt_u16_scalar_basic(uint16_t* data,
  uint32_t len, uint32_t* counters) {
  for (int i = 0; i < len; ++i) { 
    // Each bit in every input word.
    uint16_t w = data[i];
    counters[0]  += ((w >> 0)  bitand 1); 
    counters[1]  += ((w >> 1)  bitand 1);
    ...
    counters[15] += ((w >> 15) bitand 1); 
  }
}
\end{cpp}
\caption{\label{fig:pospopcnt-naive} Reference algorithm for computing the positional population count. Given some input data we compute the total number
of set bits at each position by using a branchless mask-shift-add update step.}
\end{figure}

\subsection{AVX-512}\label{sec:alg-csa}

%CSA circuits over 512-bit registers allow us to nearly solve the positional population count
%problem over 512-bit words.
We build CSA circuits over 512-bit registers using a CSA routine (Fig.~\ref{fig:csa-avx512}). The simplest network (Fig.~\ref{fig:hsillustration}) takes four 512-bit registers (or 256\,B) as inputs (Fig.~\ref{fig:pospopcnt-avx512-small}) and uses three calls to the \texttt{CSA\_AVX512} function.
We can double the size of the network and process 512\,B inputs using seven calls to the \texttt{CSA\_AVX512} function. Again, doubling the network size, we can process 1\,kB inputs using fifteen calls to the \texttt{CSA\_AVX512} function (Fig.~\ref{fig:pospopcnt-avx512}). 
Each call to the \texttt{CSA\_AVX512} function involves only two instructions (two times \texttt{vpternlogd}). Excluding load and store instructions, we use 
\begin{inparaenum}[(1)]
\item 6~instructions per block of 256\,B or
\item 14~instructions per block of 512\,B or
\item 30~instructions per block of 1\,kB. 
\end{inparaenum}
Notably, the number of instructions per input volume increases with wider circuits.

At the end of each circuit, we have a set of registers \{\texttt{B0}, \texttt{B1}, \texttt{B2}, \ldots\} corresponding, for each bit position in $i\in [0,1,\ldots, 512)$, to the sum of set bits. 
For simplicity let us assume that we are working with the smaller circuit (256\,B). Before starting the procedure, the registers \texttt{B0}, \texttt{B1}, \texttt{B2} and the 16 output bit-counters are zero-initialized. Next, the first 512\,B of inputs are process with the circuit and receive as output the updated values of \texttt{B0}, \texttt{B1}, \texttt{B2}. In addition, we also receive a new register \texttt{B3} corresponding to the fourth least significant bit of the sum of the set bits at position $i\in [0,1,\ldots, 512)$. We can update our sixteen bit-counters by checking the bits in \texttt{B3}: the first counter needs to be incremented by eight times the sum of the bits at indexes $\{0, 16, 32, \ldots, 496\}$, the second counter needs to be 
incremented by eight times the sum of the bits at indexes $\{1, 17, 33, \ldots, 497\}$, and so forth. The process of loading the next 512\,B of input data, call the circuit, and providing it with the updated values \texttt{B0}, \texttt{B1}, \texttt{B2} is repeated until no more data is available. At this point, when no more input data is available, we must increment our bit-counters with the remaining bits in \texttt{B0}, \texttt{B1}, \texttt{B2}. These are processed as \texttt{B3} except that the multiplier is 1, 2 and 4 respectively. The algorithms using the large circuits (512\,B and 1\,kB) work similarly.

The inner loop of our algorithm, where the bit-counters from the \texttt{B3} register are updated could become expensive. This loop execute 16~sums of 32~bit values and multiply the result by eight. This final multiplication can be deferred until the end of the main loop. Further, we vectorize the sum (Fig.~\ref{fig:increment-avx512}) by using sixteen vector of counters, each spanning 512~bits, or thirty-two 16-bit counters, instead of using sixteen scalar counters. Given the \texttt{B3} register, we select the least significant bit of each 16-bit word by using a mask-select operation involving the \texttt{_mm512_and_si512} intrinsic function. Next, we increment counters using a horizontal add operation involving the \texttt{_mm512_add_epi16} intrinsic. Finally, we shift each 16-bit subword of the \texttt{B3} register right by one bit with the \texttt{_mm512_srli_epi16} intrinsic function. This procedure can be written in scalar form as:\\
\begin{cpp}[0.8\columnwidth]
for (pos = 0; pos < 16; ++pos) {
  for (i = 0; i < 32; ++i) {
     counter[pos] += B[i] & 1;
     B[i] >>= 1;
  }
}
\end{cpp}
\\
with the convention that the subword $B$ at index $i$ (\texttt{B[i]}) selects a 16-bit word within \texttt{B}.
By unrolling this vectorized code, we achieve $3 \times 16=48$~instructions, without considering overhead and load/store instructions. We can add these 48~instructions to our previous counts:
\begin{inparaenum}[(1)]
\item 54 instructions per block of 256\,B or
\item 62 instructions per block of 512\,B or
\item 78~instructions per block of 1\,kB. 
\end{inparaenum}
Dividing by the number of 16-bit words processed, we get the following instruction counts per word: 
\begin{enumerate}
\item 0.42 when using blocks of 256\,B, 
\item 0.24 when using blocks of 512\,B, and 
\item 0.15 when using blocks of  1\,kB. 
\end{enumerate}
Based on these numbers, as block sizes get larger, we expect fewer instructions per input word and therefore improved performance. In comparison, we estimate about 64~instructions per 16-bit word for the scalar version. This corresponds to a $>$400-fold reduction in the number of executed instructions.

This analysis ignores the overhead cost of recovering the counts from both the vectorized counters and the running registers $\{\texttt{B0}, \texttt{B1}, \texttt{B2}, \ldots\}$. For small inputs, this overhead can occupy a dominant proportion of the total cost. For example, the sixteen vector counters span 1\,kB on their own. Hence, to achieve good performance, the input stream should at least exceed 1\,kB.

Taken together, we describe several SIMD-accelerated CSA networks of different sizes for computing the positional population count with up to $>$400-fold reduction in the number of executed instructions compared to a scalar implementation. Our SIMD-based approach can compute the positional population count for inputs that are a multiple of 256\,B. Any residual input values will be processed using a scalar subroutine.

\begin{figure}\centering
\begin{cpp}[0.8\columnwidth]
// inputs: B0, B1
//         data[i], ..., data[i + 3]
//
// temporary variables: B1A, B1B
//
// outputs: B0, B1, B2
//
CSA_AVX512(&B1A , &B0  , B0  , data[i]     , data[i + 1] );
CSA_AVX512(&B1B , &B0  , B0  , data[i + 2] , data[i + 3] );
CSA_AVX512(&B2  , &B1  , B1  , B1A         , B1B         );
\end{cpp}
\caption{\label{fig:pospopcnt-avx512-small} CSA circuit for our AVX-512 implementation in C++ processing 4~registers or 256\,B, the \texttt{CSA\_AVX512} is as in Fig.~\ref{fig:csa-avx512}.}
\end{figure}

\begin{figure}\centering
\begin{cpp}[0.8\columnwidth]
// inputs: B0, B1, B2
//         data[i], ..., data[i + 15]
//
// temporary variables: B1A, B1B, B2A, B2B,
//                      B2A, B2B
//
// outputs: B0, B1, B2, B3
//
CSA_AVX512(&B1A  , &B0  , B0  , data[i]     , data[i + 1] );
CSA_AVX512(&B1B  , &B0  , B0  , data[i + 2] , data[i + 3] );
CSA_AVX512(&B2A  , &B1  , B1  , B1A         , B1B         );
CSA_AVX512(&B1A  , &B0  , B0  , data[i + 4] , data[i + 5] );
CSA_AVX512(&B1B  , &B0  , B0  , data[i + 6] , data[i + 7] );
CSA_AVX512(&B2B  , &B1  , B1  , B1A         , B1B         );
CSA_AVX512(&B2A. , &B2  , B2  , B2A         , B2B         );
CSA_AVX512(&B1A  , &B0  , B0  , data[i + 8] , data[i + 9] );
CSA_AVX512(&B1B  , &B0  , B0  , data[i + 10], data[i + 11]);
CSA_AVX512(&B2A  , &B1  , B1  , B1A         , B1B         );
CSA_AVX512(&B1A  , &B0  , B0  , data[i + 12], data[i + 13]);
CSA_AVX512(&B1B  , &B0  , B0  , data[i + 14], data[i + 15]);
CSA_AVX512(&B2B  , &B1  , B1  , B1A         , B1B         );
CSA_AVX512(&B2B  , &B2  , B2  , B2A         , B2B         );
CSA_AVX512(&B3   , &B2  , B2  , B2A         , B2B         );
\end{cpp}
\caption{\label{fig:pospopcnt-avx512} CSA circuit for our AVX-512 implementation in C++ processing 16~registers or 1\,kB, the \texttt{CSA\_AVX512} is as in Fig.~\ref{fig:csa-avx512}.}
\end{figure}

\begin{figure}\centering
\begin{cpp}[0.8\columnwidth]
// input: B
// counter[0], ..., counter[15] are 512-bit registers
__m512i one = _mm512_set1_epi16(1); // 1-mask (000...1)
for (pos = 0; pos < 16; ++pos) { 
  __m512i masked = _mm512_and_si512(B, one)); // Select LSB (bit) with 1-mask
  counter[pos]   = _mm512_add_epi1(counter[pos], masked); // Horizontal add
  B = _mm512_srli_epi16(B, 1); // Shift 16-bit words right by 1 bit
}
\end{cpp}
\caption{\label{fig:increment-avx512} Vectorized counter increment.}
\end{figure}

\section{Experiments}\label{sec:exp}

We implemented the algorithms in C99 and we make them available online at \url{https://github.com/lemire/pospopcnt_avx512} under the Apache 2.0 license. 
Code was compiled with GCC 8.2 using the optimization flags \texttt{-O2 -march=native}. All tests were performed using a host machine with a Cannon Lake microarchitecture (Table~\ref{tab:test-cpus}). Performance was measured
using the Linux performance counters (e.g. \texttt{PERF\_COUNT\_HW\_CPU\_CYCLES} to count processor cycles). The host processor has 32\,kB of L1 cache per core, 256\,kB of L2 cache per core and 4\,MB of L3 cache. We verified that the processor is not subject to downclocking when running AVX-512 instructions\cite{inteloptimization} using the \texttt{avx-turbo}~\cite{avxturbo} benchmarking tool.
%In practice, the observed clock frequency varies slightly during our tests between 3.0\,GHz and 3.18\,GHz.

\begin{table*}
\caption{\label{tab:test-cpus} Hardware 
}
\centering
\begin{minipage}{\textwidth}
\centering
\begin{tabular}{cccccc}\toprule
Processor   & Base Frequency & Max. Frequency  & L1 data cache per core & Microarchitecture                          & Compiler\\ \midrule
Intel i3-8121U & 2.2\,GHz  & 3.2\,GHz & 32\,kB & Cannon~Lake (2018) & GCC  8.2 \\
\bottomrule
\end{tabular}
\end{minipage}
\end{table*}

For all experiments, we generate random data using a Mersenne Twister pseudorandom number generator. However, we find that performance is independent of the input data as our algorithms do not branch on the content of the data.

\begin{figure}
    \centering
    \includegraphics{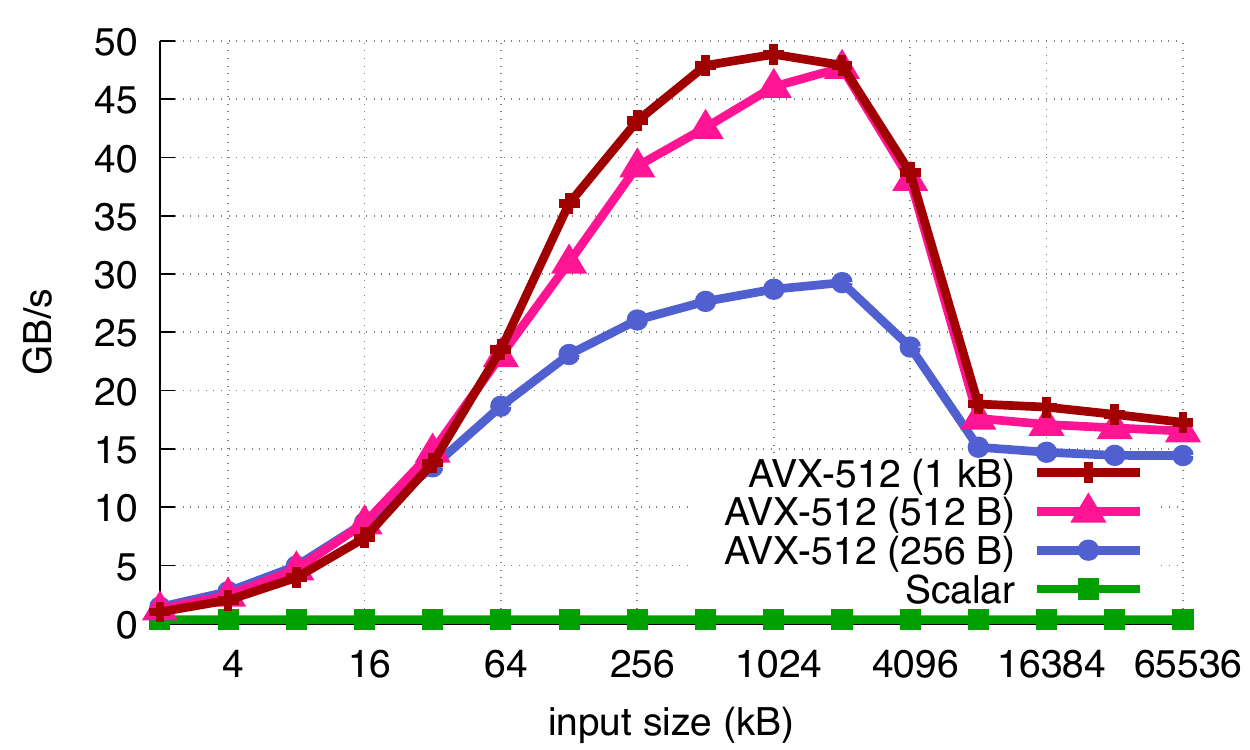}
    \caption{Processing speed (in GB/s) for various input size. Our Intel-based system has 4\,MB of L3~cache.\label{fig:speed}}
\end{figure}

Our proposed AVX-512-based approach has a fixed overhead, irrespective of the input stream size. This overhead is non-negligible for small inputs. For tiny inputs (512~words), we need about 30~instructions per 16-bit word, and almost all of these instructions are part of the overhead. However, once we reach a few thousands words (e.g., 4096~words or 8\,kB), our  AVX-512-based approach becomes beneficial. We simulated random input data for 8 and 512 kB, as well as 16, 64, 256, and 1024 MB and benchmarked our three different CSA circuits against a scalar implementation (Table~\ref{tab:results}). To guarantee reliability, we repeat each test three times and verify that the standard error is well within a 5\%. 
We also present the speeds in GB/s for input sizes up to 64\,MB (see Fig.~\ref{fig:speed}).
For the scalar algorithm, we observe a fixed speed over the range of inputs in our experiments: 17~cycles and 65~instructions per input word.  
Overall, the fastest approach is our proposed AVX-512 algorithm with large block size (1\,kB), except maybe for inputs of less than 64\,kB. Increasing the input size reduce the number of instructions per input word and simultaneously increase speed. Due to diminishing returns, the performance for 1\,GB inputs are practically identical to that of 256\,MB. In the best scenario, the AVX-512 approach uses 300~times fewer instructions compared to the scalar approach. As a reference point, the C \texttt{memcpy} function runs at a speed of 20\,GB/s for large inputs compared to 18 GB/s for the positional popcount operation on large inputs.

\begin{table}
\caption{Performance results for different input sizes: CPU cycles per 16-bit word, instructions per 16-bit word and speed in~GB/s.\label{tab:results} }\centering
\subfloat[8\,kB input]{
\begin{tabular}{lccc}\toprule
  & Cycles/word & Ins./word & Speed (GB/s) \\\midrule
Scalar  & 17 & 65 & 0.22\\
AVX-512 (256\,B) & 1.2 & 3.9 & 0.44 \\
AVX-512 (512\,B) & 1.5 & 4.7  & 0.45\\
AVX-512 (1\,kB) & 1.5 & 4.7 & 0.45\\\bottomrule
\end{tabular}
}
\subfloat[512\,kB input]{
\begin{tabular}{lccc}\toprule
  & Cycles/word & Ins./word & Speed (GB/s) \\\midrule
  & 17 & 65 & 0.36\\
& 0.22 & 0.57 & 15 \\
 & 0.15 & 0.39  & 18\\
& 0.14 & 0.29& 19\\\bottomrule
\end{tabular}
}\\
\subfloat[16\,MB input]{
\begin{tabular}{lccc}\toprule
  & Cycles/word & Ins./word & Speed (GB/s) \\\midrule
Scalar  & 17 & 65 & 0.36\\
AVX-512 (256\,B) & 0.65 & 0.52 & 9.7 \\
AVX-512 (512\,B) & 0.55 & 0.32  & 11\\
AVX-512 (1\,kB) & 0.51 & 0.23 & 12\\\bottomrule
\end{tabular}
}
\subfloat[64\,MB input]{
\begin{tabular}{lccc}\toprule
  & Cycles/word & Ins./word & Speed (GB/s) \\\midrule
  & 17 & 65 & 0.36\\
& 0.65 & 0.52 & 12 \\
 & 0.44 & 0.32  & 14\\
& 0.41 & 0.23 & 15\\\bottomrule
\end{tabular}
}\\
\subfloat[256\,MB input]{
\begin{tabular}{lccc}\toprule
  & Cycles/word & Ins./word & Speed (GB/s) \\\midrule
Scalar  & 17 & 65 & 0.36\\
AVX-512 (256\,B) & 0.47 & 0.52 & 14 \\
AVX-512 (512\,B) & 0.39 & 0.32  & 16\\
AVX-512 (1\,kB) & 0.36 & 0.22 & 18\\\bottomrule
\end{tabular}
}
\subfloat[1\,GB input]{
\begin{tabular}{lccc}\toprule
  & Cycles/word & Ins./word & Speed (GB/s) \\\midrule
  & 17 & 65 & 0.36\\
& 0.47 & 0.52 & 14 \\
 & 0.39 & 0.32  & 16\\
& 0.36 & 0.22 & 18\\\bottomrule
\end{tabular}
}
\end{table}
\section{Conclusion}\label{sec:conc}

Taken together, we introduce a novel population count operation referred to as
the \textit{positional population count} (\texttt{pospopcnt}) and introduce efficient SIMD-accelerated algorithms based on carry-saver add (CSA) circuits. We demonstrate that we can process 16-bit words at nearly the speed of a memory copy when AVX-512 is available and when the input size is sufficiently large (e.g., 256\,MB). Our best algorithm is over 40~times faster compared to a non-vectorized approach. %One limitation of our work is that the input size but must be at least a few megabytes in size for optimal results. Future work should offer fast solutions for smaller inputs.

\subsection*{Acknowledgements}

This work was funded by a Wellcome Ph.D. studentship grant 109082/Z/15/A (M.D.R.K.) as well as  by a grant (RGPIN-2017-03910) from the
Natural Sciences and Engineering (D.L.).
We are grateful to J. D. McCalpin (University of Texas at Austin) for benchmarking advice. We are grateful to members of the open-source community for reviewing and  improving the implementation.

\bibliography{fastflag}

% I am obligated to disclose this (M.D.R.K.).

%\subsection*{Author contributions statement}

%M.D.R.K conceived the project and performed experiments. M.D.R.K., W.M., and D.L. co-authored the implementation. M.D.R.K wrote the manuscript with input from W.M., and D.L. D.L. is the senior investigator. All authors reviewed the manuscript. 

%\subsection*{Competing interests}

%The authors declare no conflicts on interest.

\end{document}